\documentclass[nature,amsmath,floatfix,twocolumn,superscriptaddress]{revtex4-1}
\usepackage{amssymb,amsmath}
\usepackage{graphicx,array}
\usepackage{dcolumn}
\usepackage{psfrag}
\usepackage{bm}

\begin{document}

\title{Multiple structural transitions driven by spin-phonon couplings
  in a perovskite oxide}

\author{Claudio Cazorla}
\thanks{Corresponding Author}

\affiliation{School of Materials Science and Engineering, UNSW
  Australia, Sydney NSW 2052, Australia \\ Integrated Materials Design
  Centre, UNSW Australia, Sydney NSW 2052, Australia}

\author{Oswaldo Di\'eguez}

\affiliation{Department of Materials Science and Engineering, Faculty
  of Engineering, Tel Aviv University, IL-69978 Tel Aviv, Israel \\
  The Raymond and Beverly Sackler Center for Computational Molecular and
  Materials Science, Tel Aviv University, IL-69978 Tel Aviv, Israel}

\author{Jorge \'I\~niguez}

\affiliation{Materials Research and Technology Department, Luxembourg
  Institute of Science and Technology (LIST), 5 avenue des
  Hauts-Fourneaux, L-4362 Esch/Alzette, Luxembourg}

\maketitle

{\bf Spin-phonon interactions are central to many interesting
  phenomena, ranging from superconductivity to magnetoelectric
  effects. Yet, they are believed to have a negligible influence on
  the structural behavior of most materials. For example, magnetic
  perovskite oxides often undergo structural transitions
  accompanied by magnetic signatures whose minuteness suggests that
  the underlying spin-phonon couplings are largely irrelevant. Here we
  present an exception to this rule, showing that novel effects can
  occur as a consequence. Our first-principles calculations reveal
  that spin-phonon interactions are essential to reproduce the
  experimental observations on the phase diagram of magnetoelectric
  multiferroic BiCoO$_{3}$. Moreover, we predict that, under
  compression, these couplings lead to an unprecedented
  temperature-driven double-reentrant sequence of ferroelectric
  transitions. We propose how to modify BiCoO$_{3}$ via chemical
  doping to reproduce such striking effects at ambient conditions,
  thereby yielding useful multifunctionality.}

Most ferroelectric (FE) and ferroelastic perovskite oxides undergo
transitions involving structurally similar phases. One might guess
that spin-phonon (SP) effects should play a role in such
transformations, as it occurs in materials exhibiting more drastic
changes (e.g., Ni-based superalloys or
steal)~\cite{hickel12,fang10}. Yet, excepting the especial case of
compounds in which a magnetically-driven symmetry breaking yields
ferroelectric order~\cite{khomskii09}, SP couplings tend to have 
no impact. Even in compounds like room-temperature multiferroic
BiFeO$_{3}$ (BFO), in which SP effects affect significantly the free
energy of competing phases, their influence on the structural
transitions is minor~\cite{cazorla13}.

\begin{figure}
\centerline{
\includegraphics[width=1.00\linewidth]{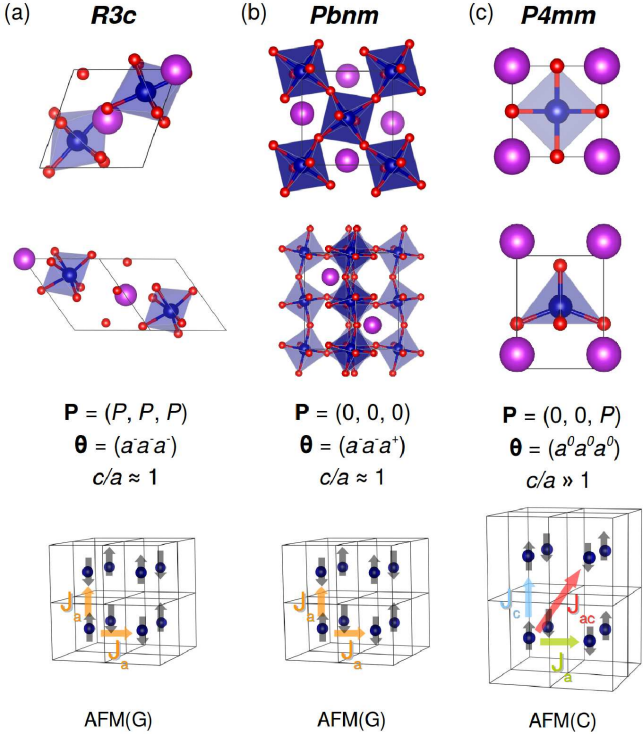}}
\caption{Structural and magnetic properties of energetically
  competitive polymorphs in bulk BiFeO$_{3}$ and BiCoO$_{3}$:
  (a)~rhombohedral $R3c$ (${\cal R}$), (b)~orthorhombic $Pbnm$ (${\cal
    O}$), and (c)~tetragonal $P4mm$ (${\cal T}$). Patterns of $O_{6}$
  octahedra rotations are expressed in Glazer's
  notation~\cite{glazer72}. The $c/a$ aspect ratio of pseudo-cubic
  lattice parameters is approximately 1 for the ${\cal R}$ and ${\cal
    O}$ phases, while it is about 1.3 for the so-called {\em
    super-tetragonal} ${\cal T}$ structure. Sketches of the
  lowest-energy spin configurations and exchange constants for a
  Heisenberg spin model of each phase are also shown.}
\label{fig1}
\end{figure}

Figure~\ref{fig1} shows the relevant polymorphs in BFO. The rhombohedral 
FE phase (${\cal R}$) that is stable at ambient conditions displays 
displacements of the Bi cations and concerted antiphase rotations of 
the O$_{6}$ octahedra about the polar [111] axis. (Axes are given in 
the pseudo-cubic setting.) There is also a paraelectric (PE) phase 
characterized by antiphase O$_{6}$ tilts about [110] and in-phase 
rotations about [001]. This orthorhombic (${\cal O}$) structure is 
stable above 1100~K~\cite{arnold09}. Magnetism in these phases is 
dominated by a strong antiferromagnetic (AFM) superexchange between 
adjacent irons, and first-principles-derived spin models yield a N{\'e}el 
temperature of about 600~K for both of them~\cite{cazorla13}. Further, 
SP couplings turn out to be very similar in both structures and thereby 
have a minute impact on their relative stability~\cite{cazorla13}.

We can conjecture that, for SP couplings to have a strong influence on
the structural transitions, the magnetic interactions in the competing
polymorphs need to be as different as possible. Interestingly,
perovskite BiCoO$_{3}$ (BCO) -- also a room-temperature multiferroic
-- complies with this requirement \cite{belik06}. At ambient
conditions BCO presents a FE tetragonal (${\cal T}$) phase with
polarization along [001] (Fig.~\ref{fig1}c) and a very distorted cell
with aspect ratio approaching 1.3. Consequently, the magnetic
interactions within a plane are stronger than across planes, rendering
a relatively low N{\'e}el temperature of about 310~K, according to our
first-principles estimate. (See Supplementary Information for more on
this result and its comparison to experiment.) At high temperatures,
BCO presents a PE ${\cal O}$ phase with $c/a \approx 1$ and a
three-dimensional spin lattice; our corresponding
first-principles-based Heisenberg model yields $T_{\rm N}
\approx$~500~K. As the spin-spin interactions, we expect the SP
couplings in BCO's ${\cal T}$ and ${\cal O}$ phases to also differ
significantly.

\begin{figure}
\centerline{
\psfrag{t}{${\cal T}$}
\psfrag{o}{${\cal O}$}
\includegraphics[width=1.00\linewidth]{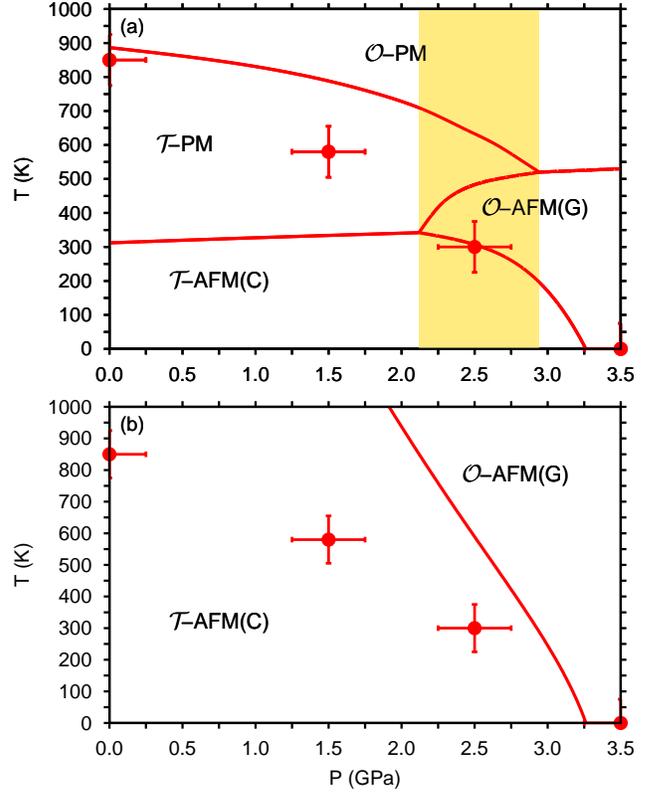}}
\caption{$P-T$ phase diagram of bulk BiCoO$_{3}$ calculated from first
  principles. (a) Spin-phonon coupling effects are considered in the
  calculation of quasi-harmonic free energies. Multiple $T$-induced
  multiferroic phase transitions occur in the region colored in
  yellow. (b) Fixed magnetic spin order (corresponding to the
  lowest-energy spin arrangement) is imposed in the calculation of
  quasi-harmonic free energies. Experimental data corresponding to the
  ferroelectric to paraelectric phase transition \cite{oka10} are
  shown for comparison (solid dots).}
\label{fig2}
\end{figure}

{\bf SP-controlled transitions at ambient pressure}

To test this, we compute the free energy of BCO's ${\cal T}$ and
${\cal O}$ phases as a function of temperature, following the
first-principles approach of Ref.~[\onlinecite{cazorla13}] (see
Methods). We obtain a critical temperature at zero pressure of
$T_{c}^{\rm the} (0) = 890(50)$~K, in agreement with the experimental
value $T_{c}^{\rm exp}(0) = 850(75)$~K \cite{oka10}
[Fig.~\ref{fig2}(a)]. Note that the ${\cal T}$--${\cal O}$ transition
occurs at a temperature at which both phases are paramagnetic (PM),
and that our method accounts for the contribution of disordered spins
to the free energy. Interestingly, if the spins are frozen in their
ground-state configuration, the transition is predicted to occur at
$2350(50)$~K, which is unrealistically high
[Fig.~\ref{fig2}(b)]. Hence, we find that magnetic disorder greatly
contributes to the stabilization of the ${\cal O}$ phase, and that SP
effects are critical to reproduce the experimental $T_{c}$.

\begin{figure}
\centerline{
\psfrag{T}{${\cal T}$}
\psfrag{O}{${\cal O}$}
\includegraphics[width=1.00\linewidth]{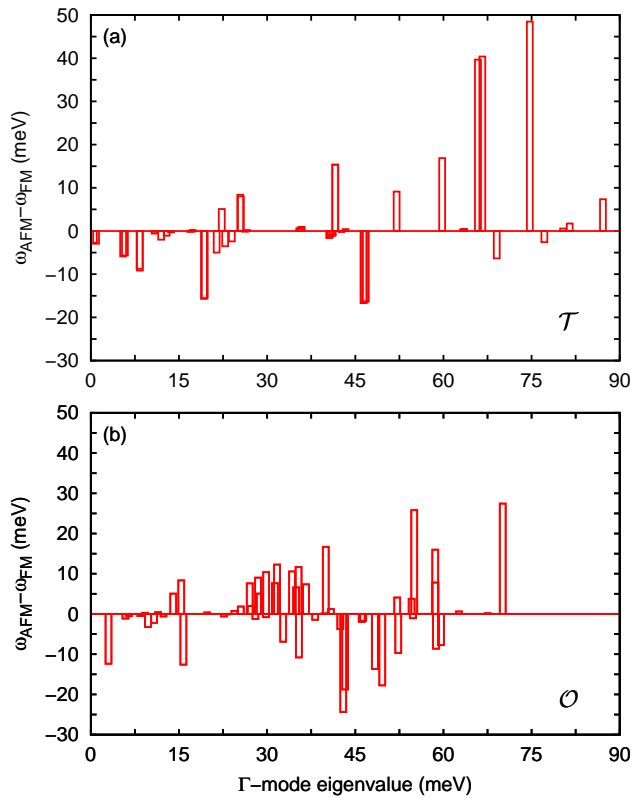}}
\caption{Analysis of spin-phonon couplings in the ${\cal T}$ and
  ${\cal O}$ phases of BCO. Vibrational frequency shifts between AFM
  and FM spin configurations calculated at the reciprocal lattice
  point $\Gamma$, $\Delta \omega = \omega_{\rm AFM} - \omega_{\rm
    FM}$, are shown as a function of eigenmode energy (where this
  energy is the one obtained for the AFM ground state). Note that
  modes with a positive (negative) shift will tend to soften (harden)
  as $T$ increases. Phonon frequency shifts for the ${\cal O}$ phase
  span over a smaller energy interval than those of the ${\cal T}$
  phase, indicating that the former structure is vibrationally softer
  than the latter. In both phases, phonon eigenmodes presenting
  largest spin couplings correspond to medium- and high-energy
  excitations dominated by Co and O atoms; in contrast, low-energy
  eigenmodes dominated by Bi and O atoms, including those associated
  to the polar distortion in BCO, present small $|\Delta \omega|$
  values.}
\label{fig3}
\end{figure}

To understand this, note how the $\Gamma$-phonon frequencies change
when considering AFM and FM (ferromagnetic) spin orders in the ${\cal
  T}$ and ${\cal O}$ phases. These frequency shifts, $\Delta \omega
\equiv \omega_{\rm AFM} - \omega_{\rm FM}$, reflect the magnitude of
SP couplings \cite{hong12} and their sign indicates which phonon
eigenmodes are more important to stabilize the corresponding PM phase
\cite{cazorla13}. Figure~\ref{fig3} shows that in the ${\cal T}$ phase
large and positive $\Delta \omega$'s mostly correspond to high-energy
phonons ($\hbar \omega \ge 60$~meV), while in the ${\cal O}$ phase
those are associated to relatively low-frequency modes ($\hbar \omega
\sim 30$~meV). Consequently, magnetic disorder favors the ${\cal O}$
polymorph. At $T_{c}^{\rm the} (0)$, for instance, fluctuating spins
provide a lattice free-energy difference of $0.168$~meV/f.u. between
${\cal T}$ and ${\cal O}$, which is three-times larger than the one
obtained when constraining AFM spin order.

{\bf SP-controlled transitions under compression, reentrant behavior}

In most perovskites, hydrostatic pressure ($P$) favors the ${\cal O}$
phase over competing polymorphs~\cite{oka10,guennou11}. Hence,
compression might help to reduce BCO's $T_{c}$ and bring it closer to
the AFM transition temperatures. To check this, we perform free-energy
calculations as a function of pressure (see Methods). Our results are
shown in Fig.~\ref{fig2}(a).

Our prediction for the ${\cal T}$--${\cal O}$ transition in the limit
of low temperatures, $P_{c}^{\rm the}(0) = 3.25(0.15)$~GPa, is in fair
agreement with the experimental value $P_{c}^{\rm exp}(0) =
3.60(0.25)$~GPa~\cite{oka10}. As regards the critical pressure and
volume drop for this transition at room temperature, we compute
$2.55(0.15)$~GPa and $\sim 11$~\%, respectively, while the
experimental values are $2.50(0.25)$~GPa and $\sim
13$~\%~\cite{oka10}. Then, as shown in Fig.~\ref{fig2}(a), the
agreement is less satisfactory at intermediate pressures. Finally, by
comparing Figs.~\ref{fig2}(a) and \ref{fig2}(b), we ratify that SP
effects are critical to reproduce the experiments.

Our phase diagram is rich in the region where structural and magnetic
transitions get close. For $P \approx 2.5$~GPa [colored area in
  Fig.~\ref{fig2}(a)], we predict that BCO presents three
temperature-driven transformations: a high-$T$ PM ${\cal O}$ phase
followed, upon cooling, by a PM ${\cal T}$ phase, a G-type AFM ${\cal
  O}$ phase, and a C-type AFM ${\cal T}$ phase. We move from a PE to a
FE phase, back to a PE structure, and finally to the FE ground
state. Note that a PE-FE-PE sequence constitutes a rare reentrant
behavior, as it is uncommon to stabilize a PE structure (typically
more disordered) by cooling down a FE phase (typically more ordered)
\cite{pociecha01,aydinol07}. Strikingly, here we find a double
reentrance, since the low-$T$ PE phase eventually transforms into the
FE ground state.

\begin{figure*}
\centerline{
\psfrag{t}{${\cal T}$}
\psfrag{o}{${\cal O}$}
\includegraphics[width=1.00\linewidth]{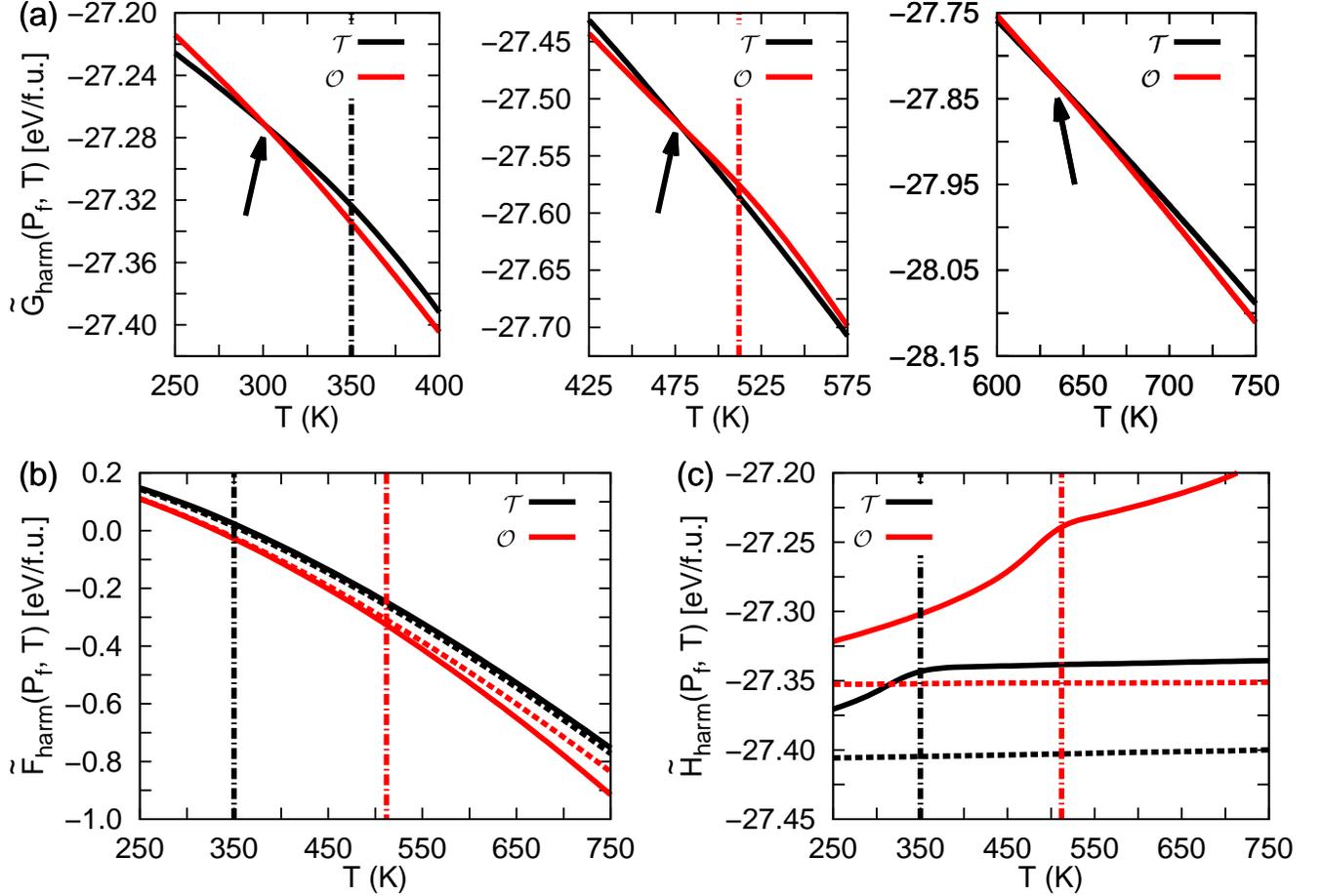}}
\caption{Calculated quasi-harmonic free energies of BCO's competing
  polymorphs, as a function of temperature and at a fixed pressure
  $P_{f} = 2.5$~GPa. (a)~Quasi-harmonic Gibbs free energy,
  $\tilde{G}_{\rm harm} \equiv \tilde{F}_{\rm harm} + \tilde{H}_{\rm
    harm}$; our estimates are accurate to within 5~meV per formula 
    unit (f.u.). (b)~Quasi-harmonic Helmholtz free energy, $\tilde{F}_{\rm
    harm} \equiv -T\tilde{S}_{\rm harm}$, where $\tilde{S}_{\rm harm}$
  represents the vibrational lattice entropy. (c)~Quasi-harmonic
  enthalpy, $\tilde{H}_{\rm harm} \equiv \tilde{E}_{0} + PV$, where $P
  \equiv -\partial (\tilde{E}_{0} + \tilde{F}_{\rm harm}) / \partial
  V$. Black arrows indicate the occurrence of structural transitions
  characterized by the thermodynamic condition $\tilde{G}_{\rm
    harm}^{\cal T} (P_{f},T_{c}) = \tilde{G}_{\rm harm}^{\cal O}
  (P_{f},T_{c})$. Black and red vertical lines signal magnetic
  spin-order transformations occurring in the ${\cal T}$ and ${\cal
    O}$ phases, respectively. Black and red dashed lines in (b) and
  (c) represent results obtained by constraining AFM magnetic spin
  order in our quasi-harmonic free-energy calculations, showing
  [panel~(b)] how spin disorder tends to stabilize the ${\cal O}$
  phase. Note that the temperature dependence of $\tilde{F}_{\rm
    harm}$ [panel~(b)] is smooth; in contrast, the slope changes in
  $\tilde{H}_{\rm harm}$ [panel~(c)], which are associated to the spin
  ordering transitions, are the main cause of the successive structural
  transformations.}
\label{fig4}
\end{figure*}

This unprecedented PE-FE-PE-FE sequence is possible because the ${\cal
  T}$ and ${\cal O}$ phases display different N\'eel temperatures and
SP couplings. In Fig.~\ref{fig4}(a) we show the temperature dependence
of the quasi-harmonic Gibbs free energy, $\tilde{G}_{\rm harm}(T)$, of
BCO's polymorphs calculated at $2.5$~GPa. We find that, whenever a
phase becomes PM, the slope of the corresponding $\tilde{G}_{\rm harm}
(T)$ curve changes noticeably; this results in three energy crossings
(structural transitions) within an interval of about 325~K. The Gibbs
free energy can be split in entropic [$\tilde{F}_{\rm harm}$,
  Fig.~\ref{fig4}(b)] and enthalpic [$\tilde{H}_{\rm harm}$,
  Fig.~\ref{fig4}(c)] terms, the latter being responsible for the
slope changes accompanying the spin transitions. Such an effect, which
is larger in the ${\cal T}$ phase, corresponds to a sizeable decrease
in the thermal expansion of the crystal when spins become disordered,
and is driven by SP couplings (the effect disappears for frozen spins;
see Supplementary Information).

{\bf Engineering multiferroic effects at ambient conditions}

The phase diagram of Fig.~\ref{fig2}(a) suggests interesting
possibilities to obtain functional properties. For example, starting
from the ${\cal O}$-AFM phase, one could use an electric field to
induce the ${\cal T}$ structure, which would result in either a loss
of spin order (if we reach the ${\cal T}$-PM phase) or a
transformation into a different AFM state (if we reach the ${\cal
  T}$-AFM phase with C-type order). For applications, one would like
to realize such phase-change effects at ambient conditions.

Chemical substitution is a practical strategy to reduce BCO's $T_{c}$
at ambient pressure. As a simple predictor for $T_{c}$, we monitor the
enthalpy difference between the ${\cal T}$ and ${\cal O}$ phases at
zero temperature, $\Delta H_{\rm eq}$, which is fast to compute from
first principles. We thus look for chemical substitutions that yield
$-0.08 \le \Delta H_{\rm eq} \le -0.03$~eV/f.u., to match the results
for pure BCO around 2.5~GPa. We find two promising cases -- namely,
BiCo$_{1/2}$Fe$_{1/2}$O$_{3}$ and Bi$_{3/4}$La$_{1/4}$CoO$_{3}$ -- for
which the enthalpy differences ($-0.075$ and $-0.033$~eV/f.u.,
respectively) lie within the targeted interval. We find both compounds
to be vibrationally stable; hence, they are good candidates to
reproduce at ambient conditions the striking effects predicted for
BCO. (See Supplementary Information for more details.)

\section*{Methods}
\paragraph{{\bf Density Functional Theory calculations.}}~We use the generalised gradient approximation to 
density functional theory proposed by Perdew, Burke, and Ernzerhof
(GGA-PBE) \cite{pbe96} as implemented in the VASP package
\cite{vasp}. We work with GGA-PBE because this is the DFT variant that
provides a more accurate description of the relative stability between
the ${\cal T}$ and ${\cal O}$ phases of BCO at zero temperature, as
discussed in the Supplementary Information. A ``Hubbard-U'' scheme
with $U = 6$~eV is employed for a better treatment of Co's $3d$
electrons \cite{dieguez11}. We use the ``projector augmented wave''
method to represent the ionic cores~[\onlinecite{bloch94}],
considering the following electrons as valence states: Co's $3p$,
$3d$, and $4s$; Bi's $5d$, $6s$, and $6p$; and O's $2s$ and $2p$. Wave
functions are represented in a plane-wave basis truncated at
$500$~eV. We use a 20-atom simulation cell that can be viewed as a $2
\times \sqrt{2} \times \sqrt{2}$ repetition of the elemental 5-atom
perovskite unit, and which is compatible with all the crystal
structures of interest here. For integrations within the Brillouin
zone (BZ), we employe $\Gamma$-centered ${\rm q}$-point grids of $6
\times 6 \times 6$. Using these parameters we obtaine enthalpy
energies converged to within $0.5$~meV per formula unit. Geometry
relaxations are performed using a conjugate-gradient algorithm that
keeps the volume of the unit cell fixed while permitting variations of
its shape and atomic positions. The relaxation stops when residual
forces fall below $0.01$~eV$\cdot$\AA$^{-1}$. Equilibrium volumes are
subsequently determined by fitting the sets of calculated energy
points to Birch-Murnaghan equations of state \cite{cazorla15}. To
treat the chemical substitutions, we work with 40-atom cell that can
be viewed as a 2$\times$2$\times$2 repetition of the elmental
perovskite cell.

\paragraph{{\bf Phonon spectrum calculations.}} The calculation of
phonon frequencies is performed with the direct method
\cite{kresse95,alfe01}, in which the force-constant matrix is
calculated in real-space by considering the proportionality between
atomic displacements and forces when the former are sufficiently
small. Large supercells need to be constructed in order to guarantee
that the elements of the force-constant matrix have all fallen off to
negligible values at their boundaries, a condition that follows from
the use of periodic boundary conditions \cite{alfe09}. Once the
force-constant matrix is calculated one can Fourier-transform it to
obtain the phonon spectrum at any ${\rm q}$-point. The impact of
long-range interactions on the calculation of long-wavelength phonons
is disregarded as we are primarily interested in the computation of
quasi-harmonic free-energies, and in such a context this factor is
known to be secondary \cite{cazorla13}. The quantities with respect to
which our phonon calculations need to be converged are the size of the
supercell, the size of the atomic displacements, and the numerical
accuracy in the sampling of the Brillouin zone. We find the following
settings to provide quasi-harmonic free energies converged to within
$5$~meV per formula unit: 160-atom supercells that can be viewed as a
$2 \times 2 \times 2$ multiple of the 20-atom unit mentioned above,
atomic displacements of $0.02$~\AA, and ${\rm q}$-point grids of $12
\times 12 \times 12$. The value of the phonon frequencies and
quasi-harmonic free energies are obtained with the PHON code developed
by Alf\`e \cite{alfe09}. In using this code we exploit the
translational invariance of the system to impose the three acoustic
branches to be exactly zero at the $\Gamma$ $q$-point, and use central
differences in the atomic forces (i.e., positive and negative atomic
displacements were considered).

\paragraph{{\bf Heisenberg model Monte Carlo simulations.}} To
simulate the effects of thermal excitations on the magnetic order of
the ${\cal T}$ and ${\cal O}$ phases, we construct several spin
Heisenberg models of the form $\hat{H} = \frac{1}{2} \sum_{ij}
J^{(0)}_{ij} S_{i}S_{j}$, in which the value of the involved exchange
constants is obtained from zero-temperature DFT calculations (see next
section and Supplementary Information). We use such models to perform
Monte Carlo (MC) simulations in a periodically-repeated simulation box
of $20 \times 20 \times 20$ spins; thermal averages are computed from
runs of $50,000$ MC sweeps after equilibration. These simulations
allow us to monitor the $T$-dependence of the magnetic order through
the computation of the AFM(C) (i.e., in the ${\cal T}$ phase) and
AFM(G) (i.e., in the ${\cal O}$ phase) order parameters, namely,
$S^{\rm C} \equiv \frac{1}{N} \sum_{i} (-1)^{n_{ix}+n_{iy}} S_{iz}$
and $S^{\rm G} \equiv \frac{1}{N} \sum_{i} (-1)^{n_{ix}+n_{iy}+n_{iz}}
S_{iz}$. Here, $n_{ix}$, $n_{iy}$, and $n_{iz}$ are the three integers
locating the $i$-th lattice cell, and $N$ is the total number of spins
in the simulation box. For the calculation of $S^{\rm C}$ and $S^{\rm
  G}$, we considered only the $z$ component of the spins because a
small symmetry-breaking magnetic anisotropy was introduced in the
Hamiltonian to facilitate the analysis (see Supplementary Information
in Ref.~[\onlinecite{escorihuela12}]).

\paragraph{{\bf Spin-phonon quasi-harmonic free-energy formalism.}} We
employ the approach described in Ref.~[\onlinecite{cazorla13}] and
generalize it to the ${\cal T}$ phase along the guidelines described
in Ref.~[\onlinecite{escorihuela12}]. In this spin-phonon free-energy
framework, the internal energy of the crystal is expressed as:
\begin{equation}
\tilde{E}_{\rm harm}(V,T) = \tilde{E}_{0}(V,T) + \frac{1}{2} \sum_{mn}
\tilde{\Xi}_{mn}(V,T) u_{m} u_{n}~,
\label{eq1}
\end{equation}
where $\tilde{E}_{0}$ represents an effective static energy,
$\tilde{\Xi}_{mn}$ an effective force constant matrix, $u$'s atomic
displacements, and the $V-T$ dependences of the various terms are
explicitly noted. The Helmoltz free energy associated to the lattice
vibrations, $\tilde{F}_{\rm harm} \equiv -T \tilde{S}_{\rm harm}$, is
calculated by finding the eigenfrequencies of the dynamical matrix
associated to $\tilde{\Xi}_{mn}$, namely, $\tilde{\omega}_{{\bf q}s}$,
and plugging them into the formula:
\begin{eqnarray}
\tilde{F}_{\rm harm} (V,T) & = & \frac{1}{N_{q}}~k_{B} T \times \nonumber \\ 
& & \sum_{{\bf q}s}\ln\left[ 2\sinh \left( \frac{\hbar \tilde{\omega}_{{\bf q}s}(V,T)}{2k_{\rm B}T} \right) \right]~,
\label{eq6}
\end{eqnarray}
where $N_{q}$ is the total number of wave vectors used for integration
in the Brillouin zone. Finally, the Gibbs free energy of each phase is
estimated as $\tilde{G}_{\rm harm} = \tilde{E}_{0} + PV +
\tilde{F}_{\rm harm}$, and the hydrostatic pressure as $P = -\partial
(\tilde{E}_{0} + \tilde{F}_{\rm harm}) / \partial V$. Our Gibbs free
energy results are accurate to within 5~meV per formula
unit. Transition points are determined via the condition
$\tilde{G}_{\rm harm}^{\cal T} (P_{c}, T_{c}) = \tilde{G}_{\rm
  harm}^{\cal O} (P_{c}, T_{c})$.

For the ${\cal O}$ phase, we showed in Ref.~[\onlinecite{cazorla13}]
that the quantities entering Eq.~(\ref{eq1}) can be calculated as:
\begin{eqnarray}
&& \tilde{E}_{0}^{\cal O}(V,T) =  E_{0}(V) + 3 \gamma_{a} (V,T) |S|^{2} J_{a}^{(0)} \, ,\\  
&& \tilde{\Xi}_{mn}^{\cal O}(V,T) =  \Phi_{mn}^{0}(V) + 6 \gamma_{a} (V,T) |S|^{2} J^{(2)}_{a, mn} \, , 
\label{eq2}
\end{eqnarray}
where $\gamma_{a} (V,T) \equiv \langle S_{i}S_{j} \rangle / |S|^{2}$
represents the correlation function between neighboring spins and
$\langle ... \rangle$ the thermal average as obtained from our MC
simulations. The rest of parameters in $\tilde{E}_{0}^{\cal O}$ and
$\tilde{\Xi}_{mn}^{\cal O}$ correspond to:
\begin{eqnarray}
&& E^{0} = \frac{1}{2}\left( E^{\rm FM}_{\rm eq} + E^{\rm G}_{\rm eq}
  \right) \, ,\\
&& \Phi_{mn}^{0} = \frac{1}{2}\left( \Phi_{mn}^{{\rm FM}} +
\Phi_{mn}^{{\rm G}} \right) \, ,\\
&& J_{a}^{(0)} = \frac{1}{6|S|^{2}}\left( E^{\rm FM}_{\rm eq} - E^{\rm
  G}_{\rm eq} \right) \, ,\\
&& J^{(2)}_{a, mn} = \frac{1}{6|S|^{2}}\left( \Phi_{mn}^{{\rm FM}} -
\Phi_{mn}^{{\rm G}} \right) \, .
\label{eq3}
\end{eqnarray}
In the equations above, superscripts ``FM'' and ``G'' indicate perfect
ferromagnetic and antiferromagnetic G-type spin arrangements,
respectively. The $J_{a}^{(0)}$ parameter describes the magnetic
interactions when the atoms remain frozen at their equilibrium
positions (see Fig.~\ref{fig1}b); typically, this captures the bulk of
the exchange couplings. Meanwhile, the $J_{a, mn}^{(2)}$ parameter
captures the dependence of the phonon spectrum on the spin
configuration (i.e., spin-phonon coupling effects).

For the ${\cal T}$ phase, we express the corresponding static energy
and force constant matrix as:
\begin{eqnarray}
&& \tilde{E}_{0}^{\cal T}(V,T) =  E_{0}(V) +  2 \gamma_{a} (V,T) |S|^{2} J_{a}^{(0)} +  \\ \nonumber
&& \gamma_{c} (V,T) |S|^{2} J_{c}^{(0)} + 4 \gamma_{ac} (V,T) |S|^{2} J_{ac}^{(0)}  \, ,\\  
&& \tilde{\Xi}_{mn}^{\cal T}(V,T) =  \Phi_{mn}^{0}(V) +  4 \gamma_{a} (V,T) |S|^{2} J^{(2)}_{a,mn} + \\ \nonumber
&& 2 \gamma_{c} (V,T) |S|^{2} J^{(2)}_{c,mn} + 8 \gamma_{ac} (V,T) |S|^{2} J^{(2)}_{ac,mn}  \, , 
\label{eq4}
\end{eqnarray}
where $\gamma_{\alpha} (V,T) \equiv \langle S_{i}S_{j} \rangle /
|S|^{2}$, with $\alpha = a, b, ac$, represent the correlation
functions between in-plane and out-of-plane neighboring spins
according to the sketch shown in Fig.~\ref{fig1}(c); the rest of
parameters in $\tilde{E}_{0}^{\cal T}$ and $\tilde{\Xi}_{mn}^{\cal T}$
can be obtained as:
\begin{eqnarray}
&& E^{0} = \frac{1}{4}\left( E^{\rm FM}_{\rm eq} + E^{\rm A}_{\rm eq} + E^{\rm C}_{\rm eq} + E^{\rm G}_{\rm eq}  \right) \, ,\\
&& \Phi_{mn}^{0} = \frac{1}{4}\left( \Phi_{mn}^{{\rm FM}} +  \Phi_{mn}^{{\rm A}} +  \Phi_{mn}^{{\rm C}} + \Phi_{mn}^{{\rm G}} \right) \, ,\\
&& J_{a}^{(0)} = \frac{1}{8|S|^{2}}\left( E^{\rm FM}_{\rm eq} + E^{\rm A}_{\rm eq} - E^{\rm C}_{\rm eq} - E^{\rm G}_{\rm eq} \right) \, ,\\
&& J^{(2)}_{a, mn} = \frac{1}{8|S|^{2}}\left( \Phi_{mn}^{{\rm FM}} + \Phi_{mn}^{{\rm A}} - \Phi_{mn}^{{\rm C}} - \Phi_{mn}^{{\rm G}} \right) \, ,\\
&& J_{c}^{(0)} = \frac{1}{4|S|^{2}}\left( E^{\rm FM}_{\rm eq} - E^{\rm A}_{\rm eq} + E^{\rm C}_{\rm eq} - E^{\rm G}_{\rm eq} \right) \, ,\\
&& J^{(2)}_{c, mn} = \frac{1}{4|S|^{2}}\left( \Phi_{mn}^{{\rm FM}}  - \Phi_{mn}^{{\rm A}} + \Phi_{mn}^{{\rm C}} - \Phi_{mn}^{{\rm G}} \right) \, ,\\
&& J_{ac}^{(0)} = \frac{1}{16|S|^{2}}\left( E^{\rm FM}_{\rm eq} - E^{\rm A}_{\rm eq} - E^{\rm C}_{\rm eq} + E^{\rm G}_{\rm eq} \right) \, ,\\
&& J^{(2)}_{ac, mn} = \frac{1}{16|S|^{2}}\left( \Phi_{mn}^{{\rm FM}} - \Phi_{mn}^{{\rm A}} - \Phi_{mn}^{{\rm C}} + \Phi_{mn}^{{\rm G}}  \right) \, .
\label{eq5}
\end{eqnarray}
In the equations above, superscripts ``FM'', ``G'', ``A'', and ``C''
indicate perfect ferromagnetic, antiferromagnetic G-type,
antiferromagnetic A-type, and antiferromagnetic C-type spin
arrangements, respectively.

\section*{Acknowledgments}
This research was supported under the Australian Research Council's Future Fellowship funding
scheme (project number FT140100135), the Israel Science Foundation through Grants 1814/14 and 
2143/14, and the Luxembourg National Research Fund through the PEARL (Grant P12/4853155 COFERMAT) 
and CORE (Grant C15/MS/10458889 NEWALLS) programs. Computational resources and technical assistance 
were provided by RES and the Australian Government through Magnus under the National Computational 
Merit Allocation Scheme.

\section*{Author contributions}
All authors contributed equally to the present work.

\section*{Additional information}
{\bf Supplementary information} accompanies this paper at xxx.
 
{\bf Competing financial interests:} The authors declare no competing financial interests.


\begin{thebibliography}{30}
\bibitem{hickel12} Hickel, T., Grabowski, B., K\"ormann, F. $\&$ Neugebauer, J. 
                   Advancing density functional theory to finite temperatures: methods and applications in steel design.
                   \textit{J. Phys. Condens. Matter} \textbf{24}, 053202 (2012).
\bibitem{fang10} Fang, C. M., Sluiter, M. H. F., van Huis, M. A., Ande, C. K. $\&$ Zandbergen, H. W.
                 Origin of predominance of cementite among iron carbides in steel at elevated temperature.
                 \textit{Phys. Rev. Lett.} \textbf{105}, 055503 (2010).
\bibitem{khomskii09} Khomskii, D. Classifying multiferroics:
  Mechanisms and effects. \textit{Physics} \textbf{2}, 20 (2009).
\bibitem{cazorla13} Cazorla, C. $\&$ ${\rm \acute{I}}$${\rm \tilde{n}}$iguez, J.
                    Insights into the phase diagram of bismuth ferrite from quasiharmonic free-energy calculations.
                    \textit{Phys. Rev. B} \textbf{88}, 214430 (2013).
\bibitem{glazer72} Glazer, A. M. The classification of tilted
         octahedra in perovskites. \textit{Acta Crystallographica Section
         B} \textbf{28}, 3384 (1972).
\bibitem{arnold09} Arnold, D. C., Knight, K. S., Morrison, F. D. $\&$ Lightfoot, P.
         Ferroelectric-paraelectric transition in BiFeO$_{3}$: crystal structure of the orthorhombic $\beta$ phase.	
         \textit{Phys. Rev. Lett.} \textbf{102}, 027602 (2009).
\bibitem{belik06} Belik, A. A., Iikubo, S., Kodama, K., Igawa, N., Shamoto, S., Niitaka, S., Azuma, M., Shimakawa, Y., 
         Takano, M., Izumi, F. $\&$ Takayama-Muromachi, E.
          Neutron powder diffraction study on the crystal and magnetic structures of BiCoO$_{3}$.
          \textit{Chem. Mater.} \textbf{18}, 798-803 (2006).
\bibitem{oka10} Oka, K., Azuma, M., Chen, W.-T., Yusa, H., Belik, A. A., Takayama-Muromachi, E., Mizumaki, M., Ishimatsu, N.,
                Hiraoka, N., Tsujimoto, M., Tucker, M. G., Attfield, J. P. $\&$ Shimakawa, Y.
                Pressure-induced spin-state transition in BiCoO$_{3}$.
                \textit{J. Am. Chem. Soc.} \textbf{132}, 9438 (2010).
\bibitem{hong12} Hong, J., Stroppa, A., ${\rm \acute{I}}$${\rm \tilde{n}}$iguez, J., Picozzi, S. $\&$ Vanderbilt, D.
                Spin-phonon coupling effects in transition-metal perovskites: a DFT+$U$ and hybrid-functional study.
                 \textit{Phys. Rev. B} \textbf{85}, 054417 (2012). 
\bibitem{guennou11} Guennou, M., Bouvier, P., Chen, C. S., Dkhil, B., Haumont, R., Garbarino, G. $\&$ Kreisel, J.  
                    Multiple high-pressure phase transitions in BiFeO$_{3}$.
                    \textit{Phys. Rev. B} \textbf{84}, 174107 (2011). 
\bibitem{pociecha01} Pociecha, D. {\sl et al}. Reentrant
         ferroelectricity in liquid crystals. \textit{Phys. Rev. Lett.}
          \textbf{86}, 3048 (2001).
\bibitem{aydinol07} Aydinol, M.K., Mantese, J.V., Alpay, S.P. A
         comparative ab initio study of the ferroelectric behaviour in
         KNO$_{3}$ and CaCO$_{3}$. \textit{J. Phys.: Condens. Matt.}
         \textbf{19}, 496210 (2007).
\bibitem{pbe96} Perdew, J. P., Burke, K. $\&$ Ernzerhof, M. 
                Generalized gradient approximation made simple.
                \textit{Phys. Rev. Lett.} \textbf{77}, 3865 (1996).
\bibitem{vasp} Kresse, G. $\&$ F\"urthmuller, J.
               Efficient iterative schemes for ab initio total-energy calculations using a plane-wave basis set. 
               \textit{Phys. Rev. B} \textbf{54}, 11169 (1996);
               Kresse, G. $\&$ Joubert, D.
               From ultrasoft pseudopotentials to the projector augmented-wave method. 
               \textit{Phys. Rev. B} \textbf{59}, 1758 (1999).
\bibitem{dieguez11} Di\'eguez, O. $\&$ ${\rm \acute{I}}$${\rm \tilde{n}}$iguez, J.
                    First-principles investigation of morphotropic transitions and phase-change functional responses 
                    in BiFeO$_{3}$-BiCoO$_{3}$ multiferroic solid solutions.
                    \textit{Phys. Rev. Lett.} \textbf{107}, 057601 (2011).
\bibitem{bloch94} Bl\"ochl P. E. 
                  Projector augmented-wave method.  
                  \textit{Phys. Rev. B} \textbf{50}, 17953 (1994).
\bibitem{cazorla15} Cazorla, C. $\&$ Boronat, J.
                    First-principles modeling of three-body interactions in highly compressed solid helium. 
                    \textit{Phys. Rev. B} \textbf{92}, 224113 (2015).
\bibitem{kresse95} Kresse, G., Furthm\"uller, J. $\&$ Hafner, J.
                   Ab initio force constant approach to phonon dispersion relations of diamond and graphite. 
                   \textit{Europhys. Lett.} \textbf{32}, 729 (1995).
\bibitem{alfe01} Alf\`e, D., Price, G. D. $\&$ Gillan, M. J.
                 Thermodynamics of hexagonal-close-packed iron under Earth’s core conditions. 
                 \textit{Phys. Rev B} \textbf{64}, 045123 (2001).
\bibitem{alfe09} Alf\`e, D. 
                 PHON: a program to calculate phonons using the small displacement method.
                 \textit{Comp. Phys. Commun.} \textbf{180}, 2622 (2009).
\bibitem{escorihuela12} Escorihuela-Sayalero, C., Di\'eguez, O. $\&$ ${\rm \acute{I}}$${\rm \tilde{n}}$iguez, J.
                        Strain engineering magnetic frustration in perovskite oxide thin films. 
                        \textit{Phys. Rev.Lett.} \textbf{109}, 247202 (2012).
\end{thebibliography}
\end{document}